# Fluidic Electrodynamics: On parallels between electromagnetic and fluidic inertia

**Alexandre A. Martins**[1]

[1]Institute for Plasmas and Nuclear Fusion & Instituto Superior Técnico
Av. Rovisco Pais, 1049-001 Lisboa, Portugal
Email: aam@ist.utl.pt

The purpose of the present work is to trace parallels between the known inertia forces in fluid dynamics with the inertia forces in electromagnetism that are known to induce resistance forces on masses both due to acceleration and at constant velocity. It is shown that the force exerted on a particle by an ideal fluid produces two effects: i) resistance to acceleration and, ii) an increase of mass with velocity. These resistance forces arise due to the fluid dragged by the particle, where the bare mass of the particle at rest changes when in motion ("dressed" particle). It is demonstrated that the vector potential created by a charged particle in motion acts as an ideal space flow that surrounds the particle. The interaction between the particle and the entrained space flow gives rise to the observed properties of inertia and the relativistic increase of mass. Parallels are made between the inertia property of matter, electromagnetism and the hydrodynamic drag in potential flow. Accordingly, in this framework the non resistance of a particle in uniform motion through an ideal fluid (D'Alembert's paradox) corresponds to Newton's first law. The law of inertia suggests that the physical vacuum can be modeled as an ideal fluid, agreeing with the space-time ideal fluid approach from general relativity.

## 1. Introduction

It has been suggested in the past that the vector potential represents some kind of fluid velocity field (Maxwell, 1861; Kirkwood, 1953; Cook, Fearn and Millonni, 1995; Marmanis, 1996, 1998; Belot 1998; Leonhardt and Piwnicki, 1999, 2000, 2001; Rousseaux and Guyon, 2002; Siegel, 2002; Martins and Pinheiro, 2009), but the consequences of this comparison have not been properly analyzed, or proved to be more than a convenient theoretical analogy. For instance, the field of Metafluids (Marmanis, 1998) makes a bridge between the equations of electromagnetism and hydrodynamics on a pure theoretical level.

In this article, we will present a different theoretical approach pointing out all the parallels between fluidic and electromagnetic inertia that will allow for the comparison between two apparently different fields. The purpose in doing so is to point leads that may help in the future understanding of the inertia property of matter.

We make use of the term "fluidic" which was introduced (with an ending in analogy with "electronic") directed mainly towards hydraulic and pneumatic control systems employing fluids instead of electrons for signal transfer and processing. Generally it involves the technique of handling fluid flows, the generation of the flow (pumps), guiding it trough conduits or channels and most importantly controlling the flow (Zimmerman, 2006). Therefore, *fluidic electrodynamics* means the generation and manipulation of the vacuum or space-time hydrodynamic flow, in accordance with the General Relativity approach of considering space-time as an ideal fluid (Grøn and Hervik, 2007), through the use of electromagnetic interactions and forces, with the aim to control and direct that flow. In the ensuing discussion, we will formulate a theoretical frame that will take us to consider the vector potential as the velocity component of a superfluid space flow, from which the inertia property can evolve.

**2. The vector potential has a space flow**

Fizeau (1851) showed experimentally for the first time that flowing water could drag light with it. He measured interference between light rays going with and against water flow, proving that the flow altered light propagation (see Figure 1(a)). The amount of light dragging by a moving medium is given by the Fresnel dragging coefficient, who predicted the effect as early as 1818. The total velocity of light $v$ in a medium of refraction index $n$ is given by:

$$v = \frac{c}{n} + \left(1 - \frac{1}{n^2}\right)u = \frac{c}{n} + \alpha u, \qquad (2.1)$$

where $c/n$ is the velocity of light in a dielectric medium and $\alpha u$ is the dragging of light by the medium with velocity $u$.

Maxwell (1861) was the first to suggest that the magnetic vector potential **A**, behaves like a moving medium, playing the velocity of a space flow around a magnetic field line (Siegel, 2002). Thus, a vector potential circular "velocity" pattern around a solenoid can be seen (compare Figure 1.(a) with Figure 1.(b)) to be equivalent to the water velocity in Fizeau's experiment of dragging light (Cook, Fearn and Millonni, 1995). The Aharonov-Bohm (AB) effect (1959) is the corresponding effect in electromagnetism consisting in the production of the phase shift between two electron waves.

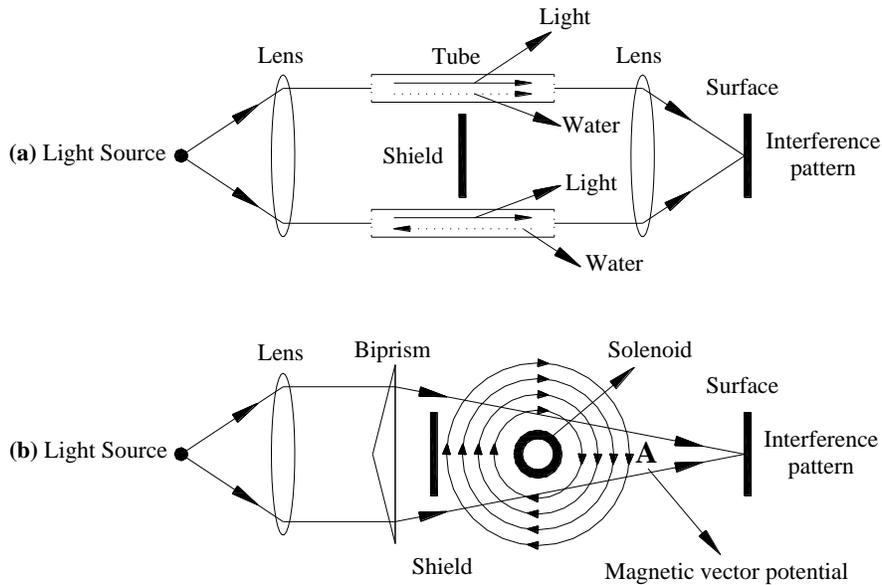

**Figure 1.** **(a)** The Fizeau Experiment, **(b)** The Optical Aharonov-Bohm Effect.

The same phenomenon of interference with light waves exists, the optical AB effect. In their interesting work on the study of light propagation in nonuniformly moving fluids, Leonhardt and Piwnicki (1999, 2000, 2001) derived the formulas for the phase change light undergoes in an optical AB effect obtained, e.g., by letting the light travel through a liquid vortex. The phase shift $\varphi_{AB}$ is given by:

$$\varphi_{AB} = 2\pi v_{AB}, \tag{2.2}$$

with,

$$v_{AB} = \frac{\omega}{c}\left(n^2 - 1\right)\frac{w}{c}, \tag{2.3}$$

where $v_{AB}$ is the light velocity, w is the vortex strength defined by $\mathbf{u}=w/r\mathbf{e}_\varphi$ in cylindrical coordinates, r is the radius at which the medium velocity is considered and $\omega$ is the optical frequency of light. The effect is small and could be enhanced by letting light take several turns inside the vortex to get cumulative effects. To this day there is not a consensus of whether light really suffers a velocity change or just simply a phase change when it passes through a physical flowing medium such as water or the vector potential of electromagnetism (Spavieri and Gillies, 2007; Boyer 2007).

The phase shift of electrons in the AB effect is considered to be a quantum topological effect where the electrons feel no force or velocity change in their path. However, there exists a competing classical explanation that sustains the phase shift to be caused by a lag effect in the transit times for the two beams which attain different velocities (in analogy with a wave from one of the beams passing through a dielectric piece and changing velocity in the process) (Liebowitz, 1965, 1966; Boyer, 1973, 1987, 2002). According to Boyer (2005, 2006, 2007), the change in velocity the electrons would suffer in the AB effect should be:

$$m\Delta\mathbf{v}_q = (q/c)\mathbf{A}_\mu\left(\mathbf{r}_q(t)\right). \tag{2.4}$$

where *m* is the mass of the particle, $\mathbf{v}_q$ is the velocity of the charge *q* and $A_\mu(r_q(t))$ is the vector potential, in the Coulomb gauge, created by the magnet, and evaluated at the position $(r_q(t))$ of the charge in motion. These velocity changes are consistent with the conservation of energy, of linear momentum and special relativity. Furthermore, this result accounts exactly for the AB phase shift as a classical electromagnetic lag effect. Preliminary experiments (Gronniger et al, 2007; Caprez, Barwick and Batelaan, 2007) have not detected a classical force on electrons passing a macroscopic solenoid. These results have been contested by Boyer (2007), thus it remains an open subject.

If it is experimentally verified that a dielectric flow (of water, for example) changes the speed of light and not just cause a phase shift, then it is reasonable to assume correspondingly that the vector potential should have a similar effect changing the velocity of light. Supporting the notion that the magnetic vector potential behaves as a superfluid space flow in the AB effect is the work of Davidowitz and Steinberg (1997) about a hydrodynamic analogy of the AB effect in superfluid helium. A moving dielectric medium appears to light as a gravitational field, modifying the spacetime metric experienced by electromagnetic waves. Light traveling through a dielectric vortex suffers an AB phase shift. On the same stance, atoms that pass through this vector potential circular flow also suffer a phase shift (Wilkens, 1994; Wei Han and Wei, 1995; Spavieri 1999).

In his fundamental article on understanding electromagnetism, Belot (1998) suggested three different interpretations of electromagnetism corresponding to three different strategies for interpreting gauge theories. In the first interpretation he considered the vector potential as a physical field on physical space in which it represents the velocity of a material ether and where the electrokinetic field, $\mathbf{E} = -\partial\mathbf{A}/\partial t$, corresponds to the acceleration of this material ether. The second interpretation is a traditional one where the electric and magnetic fields are considered to be the physically real entities. And the third interpretation is based on holonomies where the electromagnetic field is regarded as an assignment of real numbers to closed curves in space and vectors to points of space (electric field). Holonomies are used because although the vector potential, **A**, at a given point of space is not gauge invariant, the integral of **A** around a closed curve is considered to be gauge invariant.

The AB effect shows that in quantum mechanics the vector potential of electromagnetism is physically real, thus interpretation two above is false. Although Aharonov and Bohm seem to have preferred the first interpretation, Belot choose the third because of the gauge invariance in terms of holonomies, where fields are thought as properties of loops. This however destroys synchronic locality because if the field is a loop, then the state of the system depends on regions of space far away from the area of interest. This, in our view, simply does not make sense and we prefer to consider a local approach as given by the first interpretation, which is strongly supported by the fact that the vector potential has longitudinal as well as curled components (Schwab, Fuchs and Kistenmacher, 199; Rousseaux, 2003), thus showing the impossibility of interpretation three.

It is not widely known but Einstein after denying the 19$^{th}$ century concept of immobilized ether did in fact return to the ether (Einstein, 2004; Kostro, 2000; Granek, 2001) in 1916 with different and specific properties accounted for in his general theory of relativity, where he associated the metric of space with the physical properties of the vacuum, space or ether. This is compatible with fluidic electrodynamics, where the vector potential acts as a directional space "hydrodynamic" potential flow derived from charged particle movements that occur in an environment or space that behaves like a perfect fluid, much like in the way space is treated in the general theory of relativity (Grøn and Hervik, 2007).

Some authors (Rueda and Haish, 2005) suggest that inertia results from an opposing force whenever a frame accelerates relative to the vacuum. Froning (1989, 2003), Fronig and Roach (2002, 2007) relate the zero point field with fluid dynamics as having similar properties, Kelly (1976) has shown how vacuum electromagnetics can be derived exclusively from the properties of an ideal fluid and Culetu (1994), Volovik (2001), Alvarenga and Lemos (1998), Huang and Wang (2006) have treated the vacuum and cosmological phenomena as a superfluid, thus confirming the validity of our link between the vector potential induced space flow with a superfluid space flow. More recently Jacobson and Parentani (2005) have suggested that spacetime can be a kind of fluid with the properties of an ether, and Jacobson (1999) showed that particles with Planck mass form the Planck-Wheeler quantum foam of spacetime, a kind of "atoms" of spacetime forming a physical vacuum with fluid characteristics.

**3. Hydrodynamic inertia**

Newton's first principle tells us that a body remains at rest or in motion with the same speed and in the same direction unless acted upon by a force. The case of an elementary particle, a proton for instance, traveling at uniform velocity and immersed in its own vector potential flow allows us to draw some useful hydrodynamic analogies. The fact that a charged particle feels no drag when in motion with uniform velocity means that the vector potential space flow is a lossless one. This means that the particle will be subject to a perfect pressure recovery at the rear that will equal the pressure rise at the front, resulting in zero net drag. This information suggests that at uniform velocity the space flow will show no viscosity and behaves as an ideal fluid (potential flow). One finds in hydrodynamics a possible similar effect, the phenomenon of non resistance of a sphere in uniform motion through an ideal fluid, often referred to as "D'Alembert's paradox" (Faber, 2004) or "Paradox of Dirichlet" (Prandtl and Tietjens, 1934). In hydrodynamics the volume inertia force $\mathbf{f}_i$ [Nm$^{-3}$] for a particle immersed in an inviscid fluid is given by (Brennen, 1982; Moran, 1984; Faber, 2004):

$$\mathbf{f}_i = -\rho_f \mathbf{a}_f = -\rho_f \frac{d\mathbf{v}_f}{dt}. \tag{3.1}$$

Where $\mathbf{a}_f$ is the total acceleration of the fluid (Graebel, 2007). Considering that the total or convective derivative is given by:

$$\frac{d}{dt} = \frac{\partial}{\partial t} + (\mathbf{v} \cdot \nabla), \tag{3.2}$$

The volume inertia force can be written has:

$$\mathbf{f}_i = -\rho_f \frac{\partial \mathbf{v}_f}{\partial t} - \rho_f (\mathbf{v}_f \cdot \nabla) \mathbf{v}_f. \tag{3.3}$$

Where $\mathbf{v}_f$ is the fluid velocity and $\rho_f$ is the density of the hydrodynamic fluid. As is known, the first term on the right is considered to be the unsteady or temporal acceleration, responsible for the resistance of a moving body to acceleration (force of inertia), and the second term on the right is the convective acceleration, responsible for the mass increase at a given velocity. For a fluid without viscosity, an action on the body can happen only by way of pressure on its surface. A well-known result from hydrodynamic theory (Prandtl and Tietjens, 1934) shows that a particle moving in an ideal fluid will be subject to two different forces: i) the resistance to an acceleration force which is equal to the product of its mass by the acceleration (inertia force), and an ii) additional resistance force at constant velocity due to the particle mass increase ("added" mass) with the dragged fluid mass ("dressed" particle) of the surrounding fluid particles (that are dragged only when the body accelerates, and which maintain their velocity and direction when the acceleration stops) which accompany the moving body at constant velocity. According to Graebel (2007), using Newton's second law, one can write the total force needed to move a particle (to a velocity $\mathbf{v}_p$) immersed in an inviscid fluid as:

$$\mathbf{F} = (m_{body} + m_{added}) \frac{d\mathbf{v}_p}{dt} = m_{virtual} \frac{d\mathbf{v}_p}{dt}. \tag{3.4}$$

The sum of the mass of the body $m_{body}$ with the dragged or added fluid mass $m_{added}$ is sometimes called the virtual mass, which is the total "effective" mass of a body that moves immersed in an inviscid fluid. Usually the pressure distributions around bodies are given in terms of the dimensionless pressure coefficient $C_p$. The pressure coefficient (or Euler number) is defined as the static pressure divided by the dynamic pressure (White, 1988). The variation of this coefficient with velocity is approximately given by the Prandtl-Glauert rule (Anderson, 2005):

$$C_p = \frac{C_{p,0}}{\sqrt{1-M^2}}. \tag{3.5}$$

Where $C_{p,0}$ and $C_p$ are the incompressible and compressible pressure coefficients, respectively. M is the Mach number defined has:

$$M = \frac{v_f}{v_s}, \tag{3.6}$$

$v_f$ is the flow velocity of the fluid, and $v_s$ is the sound velocity in the fluid. The Prandtl-Glauert rule is a compressibility correction and is used for subsonic flow (M<1) and is generally considered to be approximately accurate for 0,3<M<0,7 (Anderson, 2005). Thus equation (3.5) is a simple but valid approximation to physical reality; it can be used to measure the resistance of the fluid on a particle with velocity $\mathbf{v}_p$, which induces a fluid velocity $\mathbf{v}_f$ around itself. Taking into account the compressible pressure corrections, the total or "effective" mass of a body that is in motion, immersed in an inviscid fluid, can be defined has:

$$m_{virtual} = \frac{m_{body}}{\sqrt{1 - v_p^2/v_s^2}} = \gamma_a \, m_{body}. \tag{3.7}$$

Where $\gamma_a$ is an acoustic gamma factor that represents the body mass increase with velocity, and the sound velocity in the fluid $v_s$ is defined has (Faber, 2004):

$$v_s = \sqrt{\frac{1}{\beta \rho_f}}, \tag{3.8}$$

$\beta$ is the compressibility of the fluid, and $\rho_f$ is the fluid density.

**4. Electrodynamic inertia**

The Euler-Lagrange electromagnetic force equation that acts on a charged particle $q$ (Semon and Taylor, 1996) is given by:

$$\frac{d}{dt}(m\mathbf{v} + q\mathbf{A}) = -q\nabla V + q\nabla_{\mathbf{A}}(\mathbf{v} \cdot \mathbf{A}). \tag{4.1}$$

Where the operator $\nabla_{\mathbf{A}}$ acts only on the magnetic vector potential $\mathbf{A}$. It has been shown (Martins and Pinheiro, 2008) that the inertia force and relativistic mass increase for a charged particle as its origin in electromagnetic forces derivable from the canonical momentum conservation:

$$\frac{d}{dt}(m\mathbf{v} + q\mathbf{A}) = 0. \tag{4.2}$$

The relativistic electromagnetic inertia force $\mathbf{F}_{iem}$ acting on the charged particle of mass m is:

$$\mathbf{F}_{iem} = \frac{d}{dt}(m\mathbf{v}) = -\frac{d(q\mathbf{A})}{dt}. \tag{4.3}$$

This identifies immediately the electromagnetic inertia force with a time change of the potential momentum $q\mathbf{A}$. Using again equation (3.2) for the total or convective derivative, and considering that the electric charge $q$ is constant, we derive the volume electromagnetic inertia force $\mathbf{f}_{iem}$:

$$\mathbf{f}_{iem} = -\rho_q \frac{\partial \mathbf{A}}{\partial t} - \rho_q (\mathbf{v} \cdot \nabla)\mathbf{A}. \tag{4.4}$$

In this way, the partial time derivative of the particle's vector potential leads to the acceleration inertia reaction force, and the space or convective derivative of the vector potential is responsible for the relativistic increase of mass of the charged particle at constant velocity. These two terms represent the global induced electric field $\mathbf{E}_i = -d\mathbf{A}/dt$ on the particle, which counteracts any imposed acceleration and induces a resistance force at constant velocity. This is illustrated in figure 2 for a positive moving charge, where $\mathbf{a}$ is the acceleration vector, $\mathbf{J}$ is the current density vector, $\mathbf{E}_i$ is the global induced electric field, and $\mathbf{F}_i$ is the force $\mathbf{E}_i$ imparts to the source particle which corresponds to the inertia reaction force, satisfying Newton's third principle. In the case of neutral matter we consider that the induced electric field created by a charge in acceleration as importance only locally for the source charge, creating inertia in the microscopic range, since on a macroscopic view the induced volume electric fields created by the positive and negative charges acceleration will have a null vector summation. On the other hand, since a neutron is composed of both positive and negative charge components (Miller, 2007), it will also possess electromagnetic inertia. In a similar way to equation (8), we can write for charged particles (with velocity $\mathbf{v}_p$):

$$\mathbf{F} = (m_{body} + m_{added})\frac{d\mathbf{v}_p}{dt} = m_{relativist}\frac{d\mathbf{v}_p}{dt}. \qquad (4.5)$$

But now the sum of the mass of the body $m_{body}$ (rest mass) with the added mass $m_{added}$ observed while in motion is called the relativistic mass, which is also the total "effective" mass of a body that is in motion, and usually defined has:

$$m_{relativist} = \frac{m_{body}}{\sqrt{1 - v_p^2/c^2}} = \gamma\, m_{body}. \qquad (4.6)$$

$\gamma$ is the Lorentz gamma factor which determines the increase of mass with velocity. The light velocity $c$ in space is defined has (Jefimenko, 1989):

$$c = \sqrt{\frac{1}{\varepsilon_0 \mu_0}}, \qquad (4.7)$$

with $\varepsilon_0$ as the permittivity of the vacuum, and $\mu_0$ as the permeability of the vacuum.

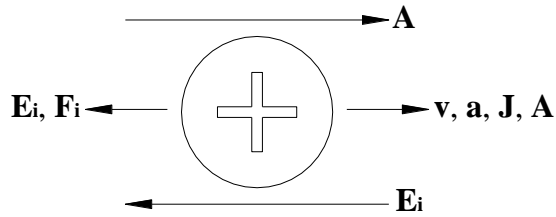

**Figure 2.** Illustration of the electromagnetic inertia mechanism for a positive moving charge.

## 5. Correspondences

We can observe a full correspondence between the hydrodynamic and relativistic Equations (3.4) (hydrodynamic inertia force) and (4.5) (relativistic inertia force), Equations (3.7) (virtual mass increase) and (4.6) (relativistic mass increase); and the hydrodynamic and electromagnetic Equations (3.8) (the sound velocity in the fluid) and (4.7) (the light velocity in space). Since the electromagnetic impedance $Z_w$ for an electromagnetic wave is given by $Z_w = \mu_0 c$ (Balanis, 1989) and the acoustic impedance $Z_a$ for an acoustic wave is given by $Z_a = \rho v_s$, then the permeability of the vacuum $\mu_0$ corresponds to the (surrounding) fluid density $\rho_f$ (Rosseaux, 2003). Comparison between equations (3.8) and (4.7) allows one to further identify the permittivity of the vacuum $\varepsilon_0$ with the compressibility of the fluid $\beta$.

Froning and Roach (2002) have compared the aerodynamic drag as given by the Prandtl-Glauert rule with the relativistic mass increase proportional to the Lorentz gamma factor and verified that generally they both have the same development below the critical velocity, respectively the sound and light velocities. We follow a different approach here in order to show that the convective derivative term of the vector potential is responsible for the relativistic increase of mass of the particle and that it has an equivalent term in hydrodynamics, responsible for the increase of mass with velocity. In both cases, the general force of inertia is given by:

$$-\rho_f \frac{\partial \mathbf{v}_f}{\partial t} - \rho_f (\mathbf{v}_f \cdot \nabla)\mathbf{v}_f \propto -\rho_q \frac{\partial \mathbf{A}}{\partial t} - \rho_q (\mathbf{v} \cdot \nabla)\mathbf{A} \qquad (5.1)$$

Equation (5.1) (in [Nm$^{-3}$]) shows the equivalence between the hydrodynamic and electromagnetic inertia terms, where we have the hydrodynamic inertia term on the left and the electromagnetic inertia term on the right. This equation shows the equivalence of both inertia forces for the general case in hydrodynamics when the fluid density is incompressible (or constant), and in the electromagnetic case when the charge density is constant. If we consider that the hydrodynamic fluid is the superfluid vacuum (Culetu, 1994; Volovik, 2001; Alvarenga and Lemos, 1998; Huang and Wang, 2006; Grøn and Hervik, 2007), we may say that the physical origin of inertia appears to come from the superfluid spacetime (which is a relativistic notion) itself, where the induced electric field $\mathbf{E}_i$, the magnetic vector potential $\mathbf{A}$, and the magnetic field are just observable properties of this quantum fluid that is everywhere. This is in accordance with Jacobson and Parentani (2005) who have suggested that spacetime can be a kind of fluid with the properties of an ether, and Jacobson (1999) which showed that particles with Planck mass form the Planck-Wheeler quantum foam of spacetime, a kind of "atoms" of spacetime forming a physical vacuum with fluid characteristics.

From equation (5.1) we can also conclude that the vacuum mass (or energy) density $\rho_f$ is affected (changed or induced) by the presence of charged particles according to $\rho_f \propto \rho_q$. In a previous article (Martins and Pinheiro, 2008) it has been demonstrated how mass has a complete electromagnetic origin, thus resolving the previous 4/3 mass paradox. Now, we can go to a more fundamental level and say that this electromagnetic mass appears to be a local vacuum energy density concentration, thus allowing us to possibly understand better the origin of inertial mass itself. This is also in accordance with Einstein's view on the interdependence of energy, fields and matter which are interchangeable with each other (Kostro, 2000). From the parallels identified in this article we can describe the correspondences put forward in Table 1.

**TABLE 1.** Comparison between hydrodynamic and electromagnetic variables.

| Hydrodynamics | Electromagnetism |
|---|---|
| Hydrodynamic fluid velocity, $\mathbf{v}_f$ | Magnetic vector potential, $\mathbf{A}$ |
| Hydrodynamic fluid acceleration ($\mathbf{a}_f$), $d\mathbf{v}_f/dt$ | Induced electric field ($\mathbf{E}_i$), $d\mathbf{A}/dt$ |
| (Induced) Mass of the fluid, $m_f$ | Electric charge, q |
| (Induced) Fluid density, $\rho_f$ | Charge density, $\rho_q$ |
| Virtual mass, $m_{virtual}$ | Relativistic mass, $m_{relativist}$ |
| Acoustic gamma factor, $\gamma_a$ | Lorentz Gamma factor, $\gamma$ |
| Acoustic impedance, $Z_a$ | Electromagnetic impedance, $Z_w$ |
| Compressibility of the fluid, $\beta$ | Permittivity of the vacuum, $\varepsilon_0$ |
| (Surrounding) Fluid density, $\rho_f$ | Permeability of the vacuum, $\mu_0$ |
| Acceleration Inertia Force, $-\rho_f(\partial\mathbf{v}_f/\partial t)$ | Acceleration Inertia Force, $-\rho_q(\partial\mathbf{A}/\partial t)$ |
| Virtual mass increase, $-\rho_f(\mathbf{v}_f \cdot \nabla)\mathbf{v}_f$ | Relativistic mass increase, $-\rho_q(\mathbf{v}\cdot\nabla)\mathbf{A}$ |
| Hydrodynamic Inertia Force, $-\rho_f(d\mathbf{v}_f/dt)$ | Electromagnetic Inertia Force, $-\rho_q(d\mathbf{A}/dt)$ |

## 6. Conclusion

We have shown the mathematical equivalence between the hydrodynamic and electromagnetic inertia force terms. Both fit perfectly into each other, indicating that the physical property of inertia can be interpreted as resulting from the interaction between moving charges and a surrounding electromagnetic spacetime ideal fluid, the vacuum energy or zero point fields, put in motion or dragged by the electric charges as represented by the magnetic vector potential.

This vacuum superfluid assumption, like that used in general relativity (Grøn and Hervik, 2007), is supported experimentally by i) the truthfulness of Newton's first law, where any particle in motion with a constant velocity will remain at that velocity and not decelerate; by ii) the experimental verification of the relativistic mass increase, that does not depend on acceleration but on velocity; and finally, on iii) the observed property of matter to always resist acceleration. All these three experimentally observed properties are consistent with our electromagnetic and hydrodynamic formulations. In this context, the vector potential can be considered as a physical field on physical space endowed with the physical property of the velocity of a kind of fluid. The vector potential can be interpreted as a polarizer of this fluid generating a space flow, due to electrical charge movements, from which the inertia property can evolve.

**References**

Aharonov Y., and Bohm, D., "Significance of electromagnetic potentials in the quantum theory," *Phys. Rev.* **115** (3), 485-491, 1959.
Alvarenga, F. G., and Lemos, N. A., "Dynamical Vacuum in Quantum Cosmology," *General Relativity and Gravitation* **30** (5), pp. 681-694, 1998.
Balanis, C. A., *Advanced engineering electromagnetics*, John Wiley & Sons, 1989.
Belot, G., "Understanding electromagnetism," *British Journal for the Philosophy of Science* **49**, pp. 531-555, 1998.
Boyer, T. H., "Classical electromagnetic deflections and lag effects associated with quantum interference pattern shifts: considerations related to the Aharonov-Bohm effect," *Phys. Rev. D* **8**, pp. 1679-1693, 1973.
Boyer, T. H., "The Aharonov-Bohm effect as a classical electromagnetic lag effect: an electrostatic analogue and possible experimental test," *Il Nuovo Cimento* **100B**, pp.685-701, 1987.
Boyer, T. H., "Semiclassical explanation of the Matteuci-Pozzi and Aharonov Bohm phase shifts," *Found. Phys.* **32**, pp.41-49, 2002.
Boyer, T. H., "The paradoxical forces for the classical electromagnetic lag associated with the Aharonov-Bohm phase shift," (2005), arXiv:physics/0506180v1.
Boyer, T. H., "Proposed experimental test for the paradoxical forces associated with the Aharonov-Bohm phase shift," *Found. Phys. Lett.* **19** (5), 491-498 (2006).
Boyer, T. H., "Unresolved classical electromagnetic aspects of the Aharonov-Bohm phase shift," (2007), arXiv:0709.0661v1.
Brennen, C. E., "A review of added mass and fluid inertial forces," Report Number CR 82.010, Naval Civil Engineering Laboratory - USA, 1982.
Caprez, A., Barwick, B., and Batelaan, H., "Macroscopic test of the Aharonov-Bohm effect," *Phys. Rev. Lett.* **99**, 210401, 2007.
Cook, R. J., Fearn, H., and Millonni, P. W., "Fizeau's experiment and the Aharonov–Bohm effect," *American Journal of Physics* **63** (8), 1995.
Culetu, H., "A "conformal" perfect fluid in the classical vacuum," *General Relativity and Gravitation* **23** (3), pp. 283-290, 1994.
Davidowitz, H., and Steinberg, V., "On an analog of the Aharonov-Bohm effect in superfluid helium," *Europhys. Lett.* **38** (4), pp. 297-300, 1997.
Einstein, A., "*Sidelights on Relativity – Ether and the Theory of Relativity + Geometry and Experience*," 1922, Elegant Ebooks, 2004.
Faber, T. E., *Fluid dynamics for physicists*, Cambridge University Press, 2004.
Fizeau, M. H., "Sur les hypothèses relatives à l'éther lumineux, et sur une expérience qui paraît démontrer que le mouvement des corps change la vitesse avec laquelle la lumière se propage dans leur intérieur," *Comptes. Rendus de l' Academie des Science de Paris* **33**, pp. 349-355, 1851.
Fresnel, A. J., "Lettre de M. Fresnel à M. Arago, sur l'influence du movement terrestre dans quelques phénomenes d' optíque," *Ann. de Chimie et Physique* **9** (57), 1818.
Froning, H. D., "Application of Fluid Dynamics to the Problems of Field Propulsion and Ultra High-Speed Flight," *American Institute of Aeronautics and Astronautics* AIAA-90-0563, 1989.

Froning, H.D., and Roach, R.L., "Preliminary simulations of vehicle interactions with the quantum vacuum by fluid dynamic approximations" in proceedings of 38th *AIAA/ASME/SAE/ASEE Joint Propulsion Conference*, AIAA-2002-3925, 2002.

Froning, H. D., "The "Quantum Interstellar Ramjet" Revisited," *American Institute of Aeronautics and Astronautics* AIAA-2003-4883, 2003.

Froning, H. D., and Roach, R. L., "Fluid Dynamic Simulations of Warp Drive Flight Through Negative Pressure Zero-Point Vacuum," in proceedings of *Space Technology and Applications International Forum* (*STAIF*-2007), edited by M. S. El-Genk, AIP Conference Proceedings 880, New York, 2007, pp. 1125-1131.

Graebel, W. P., *Advanced fluid mechanics*, Academic Press, 2007.

Granek, G., "Einstein's Ether: Why did Einstein Come Back to the Ether?," *Apeiron* **8** (3), 2001.

Grøn, ∅., and Hervik, S., *Einstein's general theory of relativity: With modern applications in cosmology*, Springer, 2007.

Gronniger, G., Simmons, Z., Gilbert, S., Caprez, A., Batelann, H., "The Aharonov-Bohm effect, phase or force," (2007), http://www.unl.edu/hbatelaan/pdf/ABpaperv3.pdf, accessed September 17, 2007.

Huang, W.-H., and Wang, I-C., "Quantum Perfect-Fluid Kaluza-Klein Cosmology," *International Journal of Modern Physics A* **21** (22), pp.4463-4478, 2006.

Jacobson, T. A., "Trans-Planckian Redshifts and the Substance of the Space-Time River," *Progress of Theoretical Physics Supplement*, No. **136** (1), 1999.

Jacobson, T. A., and Parentani, R., "An Echo of Black Holes," *Scientific American*, December 2005.

Jefimenko, O. D., *Electricity and magnetism*, 2$^{nd}$ ed., Electret Scientific Co., Star City, 1989.

Kelly, E.M., "Vacuum electromagnetics derived exclusively from the properties of an ideal fluid," *Nuovo Cimento* **32B** No 1, pp. 117-137, 1976.

Kostro, L., *Einstein and the Ether*, Apeiron 2000.

Leonhardt, U., and Piwnicki, P., "Optics of nonuniformly moving media," *Phys. Rev. A* **60**, pp. 4301-4312, 1999.

Leonhardt, U., and Piwnicki, P., "Light in Moving Media," *Contemporary Physics* **41** (5), pp. 301-308, 2000.

Leonhardt, U., and Piwnicki, P., "Optics of Moving Media," *Applied Physics B* **72**, pp. 51-59, 2001.

Liebowitz, B., "Significance of the Aharonov-Bohm effect," *Nuovo Cimento* **38**, pp. 932-950, 1965.

Liebowitz, B., "Significance of the Aharonov-Bohm effect. Rebuttal of a criticism," *Nuovo Cimento* **46B**, pp. 125-127, 1966.

Marmanis, H., "On the analogy between electromagnetism and turbulent hydrodynamics," arXiv:hep-th/9602081v1, pp. 1-14, 1996.

Marmanis, H., "Analogy between the Navier-Stokes equations and Maxwell's equations: Application to turbulence," *Physics of Fluids* **10** (6), pp.1428-1437, 1998.

Martins, A. A., and Pinheiro, M. J., "On the electromagnetic origin of inertia and inertial mass," *Int. J. Theor. Phys.*, **47** (10), pp. 2706-2715, 2008.

Martins, A. A., and Pinheiro, M. J., "Fluidic Electrodynamics: Approach to electromagnetic propulsion," *Physics of Fluids* **21**, 097103, 2009.

Maxwell, J. C., "On Physical Lines of Force, Part I: The theory of molecular vortices applied to magnetic phenomena," *Philosophical Magazine and Journal of Science*, **21** No. 139, pp. 161-175, March 1861.

Miller, G. A., "Charge densities of the neutron and proton," *Phys. Rev. Lett.* **99**, 112001, 2007.

Moran, J., *An introduction to theoretical and computational aerodynamics*, Dover Publications, 1984.

Prandtl, L., and Tietjens, O, G., *Applied hydro and aeromechanics*, Dover Publications, New York, 1934.

Rousseaux, G., and Guyon, É., "À propos d'une analogie entre la mécanique des fluides et l' électromagnétisme," *Bulletin de l'Union des Physiciens*, **96**, pp. 107-136, 2002.

Rousseaux, G., "On the physical meaning of the gauge conditions of classical electromagnetism: the hydrodynamics analogue viewpoint," *Annales de la Fondation Louis de Broglie* **28** (2), pp. 261-270, 2003.

Rueda, A., and Haish, B., "Gravity and the quantum vacuum inertia hypothesis," *Ann. Phys.* (Leipzig) **14**, No. 8, pp. 479-498, 2005.

Schwab, A., Fuchs, C., and Kistenmacher, P., "Semantics of the irrotational component of the magnetic vector potential, A," *IEEE Antennas and Propagation Magazine* **39** (1), pp. 46-51, 1997.

Semon, M. D., and Taylor, J. R., "Thoughts on the magnetic vector potential," *American Journal of Physics*, **64** (11), pp. 1361-1369, 1996.

Siegel, D. M., *Innovation in Maxwell's electromagnetic theory: Molecular vortices, displacement current, and light*, Cambridge University Press, 2002.

Spavieri, G., "Quantum Effect of the Aharonov-Bohm Type for Particles with an Electric Dipole Moment," *Phys. Rev. Lett.* **82** (20), pp. 3932-3935, 1999.

Spavieri, G., and Gillies, G. T., "A Test of the Fizeau Type for the Magnetic Model of Light in Moving Media," *Chinese Journal of Physics* **45** (1), pp.12-31, 2007.

Volovik, G.E., "Superfluid analogies of cosmological phenomena," *Physics Reports* **351** (4), pp. 195-348, 2001.

Wilkens, M., "Quantum phase of a moving dipole," *Phys. Rev. Lett.* **72** (1), pp.5-8, 1994.

Wei, H., Han, R., and Wei, X., "Quantum Phase of Induced Dipoles Moving in a Magnetic Field," *Phys. Rev. Lett.* **75** (11), pp. 2071-2073, 1995.

White, F. M., *Fluid Mechanics*, 2th edition, McGraw-Hill International Editions, 1988.

Zimmerman, W. B., *Microfluidics: History, theory and applications*, SpringerWien New York, 2006.